\documentclass[12pt,preprint,usenatbib]{aastex}
\bibpunct{(}{)}{;}{n}{}{,}
\usepackage{graphicx}   
\usepackage{amssymb}   
\DeclareGraphicsExtensions{.ps,.pdf,.png}
\makeatletter
\def\fps@figure{htbp}
\makeatother
\def \aap  {A\&A}

\def \apj  {ApJ}
\def \apjs  {ApJS}

\def \mnras {MNRAS}
\def \nat {Nature}
\def \apjl {ApJL}

\def\simgeq{\mbox{\raisebox{-1.0ex}{$\stackrel{>}{\sim}$}}}
\def\paren#1{\left( #1 \right)}
\def\bra#1{\left[ #1 \right]}

\def\ltsima{$\; \buildrel < \over \sim \;$}
\def\lsim{\lower.5ex\hbox{\ltsima}}
\def\gtsima{$\; \buildrel > \over \sim \;$}
\def\gsim{\lower.5ex\hbox{\gtsima}}

\begin{document}
\title{Magnetization Degree of Gamma-Ray Burst Fireballs:
Numerical Study}
\author{Richard Harrison\altaffilmark{1\star} and Shiho Kobayashi\altaffilmark{1}}  
\altaffiltext{1}{Astrophysics Research Institute, Liverpool John Moores University,
	IC2, Liverpool Science Park, Liverpool, L3 5RF}
\altaffiltext{$\star$}{R.M.Harrison@2006.ljmu.ac.uk}
%\date{Accepted ??. Received ??; in original form ??}
%\pagerange{\pageref{firstpage}--\pageref{lastpage}} \pubyear{2012}
%\maketitle
\label{firstpage}

\begin{abstract}
 The relative strength between forward and reverse shock emission
 in early gamma-ray burst afterglow reflects that of magnetic
 energy densities in the two shock regions.  We numerically show that
 with the current standard treatment, the fireball magnetization is
 underestimated by up to two orders of magnitude. This discrepancy  
 is especially large in the sub-relativistic reverse shock regime
 (i.e. the thin shell and intermediate regime) where most
 optical flashes were detected. We provide new analytic estimates of the
 reverse shock emission based on a better shock approximation,
 which  well describe numerical results in the intermediate regime. 
 We show that the reverse shock temperature at the onset of afterglow is
 constant, $(\bar{\Gamma}_d-1)\sim 8\times10^{-2}$, 
 when the dimensionless parameter $\xi_{0}$ is more than several.  
 Our approach is applied to case studies of GRB 990123 and 090102, and
 we find that magnetic fields in the fireballs are even stronger than
 previously believed. However, these events are still likely to be due
 to a baryonic jet with $\sigma \sim 10^{-3}$ for GRB 990123 and 
 $\sim 3\times10^{-4}-3$ for GRB 090102.
\end{abstract}
\keywords{
gamma rays: bursts --- magnetic fields --- hydrodynamics --- ISM: jets and outflows}
 
%%%%%%%%%%%%%%%%%%%%%%%%%%%%%%%%%%%%%%%%%%%%%%%%%%%%%%%%%%%%%%%%%%%%%%%%%%%%%%%%%%%%%%%%%%%%%%%%%%%%

\section{INTRODUCTION}
\label{sec:intro}

A widely accepted model for producing gamma-ray bursts (GRBs) is based
on the dissipation of a relativistic outflow (e.g. Piran 2004; Zhang \&
M\'{e}sz\'{a}ros 2004). The internal energy produced by shocks is
believed to be radiated via synchrotron emission. Although the presence
of strong magnetic fields is crucial in the model, the origin and its
role in the dynamics are still unknown. Understanding the nature of the
relativistic outflow, especially the energy content, acceleration and
collimation, is a major focus of international theoretical and
observational efforts. Relativistic outflow from a GRB central engine is
conventionally assumed to be a baryonic jet, producing synchrotron
emission with tangled  magnetic fields generated locally by
instabilities in shocks (Medvedev \& Loeb 1999; Nishikawa et al. 2005;
Spitkovsky 2008). Recently an alternative
magnetic model is attracting attention from researchers (e.g. 
Drenkhahn \& Spruit 2002; Fan et al. 2004;
Zhang \& Kobayashi 2005; Lyutikov 2006; Giannios 2008; Mimica et
al. 2009, 2010; 
Zhang \& Pe'er 2009; Zhang \& Yan 2011; Narayan et al. 2011; Granot 2012). 
The rotation of a black hole and an accretion disk might cause a helical
outgoing Magnetohydrodynamic (MHD) wave which accelerates material frozen into
the field lines (Tchekhovskoy et al. 2008; McKinney \& Blandford 2009; 
Komissarov et al. 2009). 
In the magnetic model, a fireball is expected to be endowed with
primordial magnetic fields from the central engine. 

The first detection of ten percent polarization of an optical afterglow
just 160~sec after the GRB explosion (Steele et al. 2009) opens the exciting possibility of directly measuring
the magnetic properties of the GRB outflow. Recently polarization
measurements of the prompt gamma-ray emission were also reported 
(Kalemci et al. 2007; McGlynn et al. 2007; G{\"o}tz et al. 2009; Yonetoku et
al. 2011). Although these polarization measurements suggest that at least 
some GRB outflows contain ordered magnetic fields and they are still
baryonic, the sample is small and further observations will be
necessary to confirm the magnetic model and/or to understand the
role of magnetic fields in the dynamics. In this paper, we revisit the
magnetization estimate of the GRB outflow (hereafter ``fireball'')
based on photometric observations of early optical
afterglow. It is more sensitive to the magnetic energy density, 
rather than the length scale of magnetic fields in the fireball, 
and it is complementary to polarimetric methods (e.g. Lazzati 2006; Toma
et al. 2009).  

A steep decay in early optical afterglow light curves is usually
considered as a signature of the reverse shock emission (e.g. Akerlof et
al.1999; Sari \& Piran 1999; Meszaros \& Rees 1999; Soderberg \&
Ramirez-Ruiz 2002; Li et al. 2003; Fox et al. 2003; Nakar \& Piran
2005).  The early emission contains precious  
information on the original ejecta from the central engine. The
magnetization of the fireball can be evaluated by using the relative
strength of the forward and reverse shock  emission (Fan et al. 2002, 2005; 
Zhang et al. 2003; Kumar \& Panaitescu 2003; Gomboc et al. 2008). However,
the standard  method is obtained by using  a simplified shock
dynamics model, and Nakar \& Piran (2004) have shown that the simple model 
is inaccurate in the intermediate regime between the thin and thick
shell extremes. Since most observed events are in the intermediate regime,
here we numerically re-examine the interplay between the forward and reverse
shock emission at the onset of afterglow. In Section \ref{sec:FSRS} we
set out a simple conventional approach to understanding
the two shock emissions and refine the definition of the
magnetization parameter in Section \ref{sec:Mag}. In Section
\ref{sec:inter} we consider a  new approximation to  discuss
the reverse shock emission in the intermediate regime. 
In Section \ref{sec:Method}, we then test these analytic approximations
with numerical simulations. In Section \ref{sec:case} we present case 
studies of GRB 990123 and 090102 in terms of the magnetization
parameter. Finally in Section \ref{sec:conclusion} we summarize the
results.

%%%%%%%%%%%%%%%%%%%%%%%%%%%%%%%%%%%%%%%%%%%%%%%%%%%%%%%%%%%%%%%%%%%%%%%%%%%%%%%%%%%%%%%%%%%%%%%%%%%%
\section{Forward and Reverse Shock}
\label{sec:FSRS}
We consider a homogeneous fireball 
\footnote{
Since we assume magnetic fields in the fireball do not affect the
reverse shock dynamics, our magnetization estimates are valid only when the 
fireball is weakly magnetized. The model consistency will be checked later 
when our results are applied to specific events (see Section 6).
Because of the relativistic beaming effect, the radiation from a jet
before the jet break can be described by a spherical model.}
of energy $E$ and a baryonic load of
total mass $M$ confined initially in a sphere of radius $r_0$. 
We define the dimensionless entropy $\Gamma_0 \equiv E/Mc^2 \gg 1$. 
This fireball expands into a  homogeneous interstellar medium (ISM) of 
particle density $n_1$. This can be considered to be a free expansion in
its initial stage. After a short acceleration phase, the motion becomes
highly relativistic,  and a narrow shell is formed. After the fireball shell
uses up all its internal energy,  it coasts with a Lorentz factor of
$\Gamma_0$ and the radial width $\Delta_0 \sim r_0$. 

The deceleration process of the shell is described with two shocks: a
forward shock propagating into the ISM and a reverse shock 
propagating into the shell. The forward shock is always ultra
relativistic, while the evolution of the reverse shock is determined by
a dimensionless parameter $\xi_0=(l/\Delta_0)^{1/2}\Gamma_0^{-4/3}$ where
$l=(3E/4\pi m_pn_1c^2)^{1/3}$ is the Sedov length and $m_p$ is the 
proton mass. If $\xi_0 < 1$ (so called thick shell
case),  the reverse shock becomes relativistic in the frame of 
the unshocked shell material and it drastically decelerates the
shell. If $\xi_0 > 1$ (thin shell case), the reverse shock is inefficient at slowing down
the shell. The deceleration radius $r_d$ and the  Lorentz factor
$\Gamma_d$ of the shocked material at $r_d$ are usually approximated as 
$r_{d} \sim l^{3/4}\Delta_0^{1/4}$ and $\Gamma_d \sim (l/\Delta_0)^{3/8}$
for $\xi_0 < 1$, and $r_{d} \sim l/\Gamma_0^{2/3}$ and $\Gamma_d \sim \Gamma_0$ 
for $\xi_0 > 1$ (Sari and Piran 1995; Kobayashi et al. 1999). 
After the deceleration, the profile of the shocked ISM medium begins to
approach the Blandford \& McKee (1977) solution. 

We first discuss the forward and reverse shock emission by using these
conventional estimates, and the accuracy (i.e. correction factors) will
be numerically examined later. The deceleration of a shell happens at an
observer time  
\begin{equation}
t_{d} = C_t \frac{l}{c\Gamma_0^{8/3}},
\label{eq:td}
\end{equation}
where $C_t \sim (\Gamma_d/\Gamma_0)^{-8/3} \sim \max(1,\xi_0^{-2})$ and 
all the correction factors in this paper, including $C_t$, are defined as
ones relative to the conventional thin shell estimates. 
At the deceleration time,  the forward and reverse shock regions 
have almost the same Lorentz factor and internal energy density.
However, the reverse shock region has a much larger mass density 
and therefore
it has a lower temperature.  Introducing a magnetization parameter
$R_B\equiv\epsilon_{B,r}/\epsilon_{B,f}$, it is shown that the
typical frequencies $\nu_m$ and peak fluxes $F_{\nu,max}$ 
of the synchrotron emissions from the two shocks are related as 
(Kobayashi \& Zhang 2003; Zhang et al. 2003)
\begin{equation}
\frac{\nu_{m,r}(t_d)}{\nu_{m,f}(t_d)} =  C_m \Gamma_0^{-2} R_B^{1/2},  
\ \  \ \ 
\frac{F_{\nu,max,r}(t_d)}{F_{\nu,max,f}} = C_F \Gamma_0 R_B^{1/2},
\label{eq:relation} 
\end{equation}
where $C_m \sim (\Gamma_d/\Gamma_0)^{-4} \sim
\max(1,\xi_0^{-3})$ and $C_F \sim (\Gamma_d/\Gamma_0)^2 \sim
\min(1,\xi_0^{3/2})$ are correction factors,
the subscripts $r$ and $f$ indicate reverse and forward shock, 
respectively. We have assumed that the electron equipartition parameter 
$\epsilon_e$ and the electron power-law index $p$ 
are the same for the two shock regions, but with different magnetic 
equipartition parameter $\epsilon_B$ as parametrized by $R_B$. The
reason we introduce the $R_B$ parameter is that the fireball might be
endowed with primordial magnetic fields from the central engine. 
We can give a simple relation $\nu_{c,r}/\nu_{c,f}\sim R_B^{-3/2}$ between the
cooling break frequencies of the two shock emissions (Zhang et
al. 2003). As we will see in the next section, this simple estimate is
good enough for the magnetization estimate.     

\section{Magnetization Estimates}
\label{sec:Mag}

\begin{figure}
\begin{center}
\includegraphics[height=16cm]{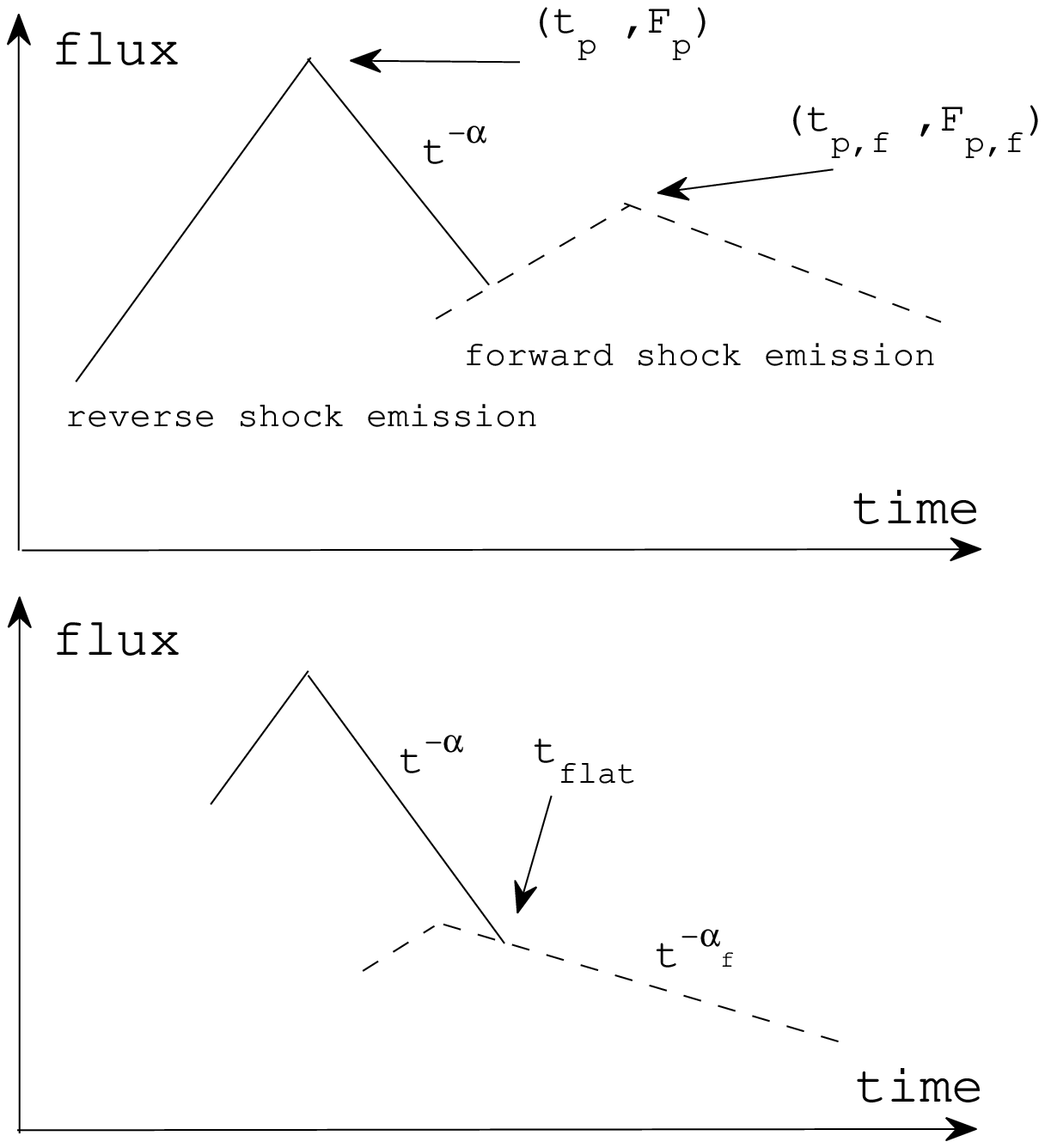}
\caption{Early steep decay: optical afterglow, produced as a composition of 
the reverse shock emission (solid line) and forward shock emission
 (dashed line). 
Two peaks (top panel) and a flattening (bottom panel) in the light curve.
\label{fig:twopeaks}}
\end{center}
\end{figure}

For no or moderate primordial magnetic fields in a fireball 
\footnote{ Even if $\epsilon_B$ at the forward shock is very low 
(e.g. $\epsilon_{B,f}\sim 10^{-5}$; Kumar \& Barniol Duran 2009), we 
expect the typical frequency of the reverse shock emission is much lower 
than that of the forward shock emission 
$\nu_{m,r} \ll \nu_{m,f}$.  For typical events ($\Gamma_{0} > 10^{2}$
and $C_m\sim 1$), extreme magnetization $R_B \sim \Gamma_{0}^4 > 10^8$
is needed to achieve $\nu_{m,r} \sim \nu_{m,f}$ at the peak time.},
we expect $\nu_{m,r} \ll \nu_{m,f}$ and $\nu_{c,r}\lsim \nu_{c,f}$
at the peak time of the reverse shock emission $t_p\sim t_d$. 
The optical band $\nu_{opt}$ should satisfy a relation
$\nu_{m,r} < \nu_{opt} < \nu_{c,r}$ during the early steep decay phase 
of the reverse shock emission. Otherwise the decay is much slower or
faster than  the typical decay $t^{-2}$ 
(Kobayashi 2000). 
There are four possible relations between the break frequencies and the
optical band 
at the peak time $t_p$:
(a) $\nu_{m,r}<\nu_{opt}<\nu_{m,f}<\nu_{c,r} < \nu_{c,f}$, 
(b) $\nu_{m,r}<\nu_{opt}<\nu_{c,r}<\nu_{m,f} < \nu_{c,f}$, 
(c) $\nu_{m,r}<\nu_{opt}<\nu_{c,r}<\nu_{c,f} < \nu_{m,f}$, and
(d) $\nu_{m,r}<\nu_{m,f}<\nu_{opt}<\nu_{c,r} < \nu_{c,f}$. 
In the cases (a) and (b), the forward shock emission
peaks at $t_{p,f}$ when the typical frequency $\nu_{m,f}$ goes through the
optical band (the top panel in figure \ref{fig:twopeaks}). Using  
$\nu_{m,f}\propto t^{-3/2}$, we get the peak time
and peak flux ratio 
\begin{eqnarray}
R_t&\equiv& t_{p,f}/t_{p}=\paren{\nu_{m,f}(t_{p})/\nu_{opt}}^{2/3},
\label{eq:Rt}\\
R_F&\equiv&  
F_{p}/F_{p,f} =(F_{\nu,max,r}(t_{p})/F_{\nu,max,f})(\nu_{opt}/\nu_{m,r}(t_{p}))^{-(p-1)/2}
\label{eq:Rf}
\end{eqnarray}
where $F_{p}$ and $F_{p,f}$ are optical peaks in the time domain, while 
$F_{\nu,max,r}$ and $F_{\nu,max,f}$
are peaks in the spectral domain for a given time. The hydrodynamics
evolution of a reverse shocked shell is investigated in Kobayashi \& Sari
(2000), and the decay index $\alpha \sim (3p+1)/4 \sim 2$ of the reverse shock
emission is found to be almost independent of $\xi_0$ 
when $\nu_{m,r}<\nu_{opt} < \nu_{c,r}$. Combining Equations
(\ref{eq:relation}), (\ref{eq:Rt}) and (\ref{eq:Rf}), 
we obtain (Gomboc et al. 2008)
\begin{equation}
R_B=\paren{\frac{R_F^3\Gamma_0^{4\alpha-7}}
{C_F^3C_m^{2(\alpha-1)}R_t^{3(\alpha-1)}}}^{2/(2\alpha+1)}
\sim \paren{\frac{R_F^3\Gamma_0}{C_F^3C_m^2R_t^3}}^{2/5}.
\label{eq:RB}
\end{equation}
At this stage, we assume that $\Gamma_0$ is a known quantity, and 
we will discuss how to estimate $\Gamma_0$ from early afterglow 
observations in section 5.3. In the case (c), if the forward 
shock emission makes a transition from the fast cooling to 
the slow cooling regime before it peaks, it becomes equivalent to 
the case (b). The estimate (\ref{eq:RB}) is valid. On the 
other hand, if it is still in the fast cooling regime when 
$\nu_{c,f}$ crosses the optical band, the forward shock emission
rises and decays very slowly as $t^{1/6}$ and
$t^{-1/4}$ (Sari et al. 1998). Since this behavior is not consistent with 
most early afterglows, we do not discuss the details
\footnote{In this case, we need an additional relation 
$\nu_{m,f}(t_p)/\nu_{c,f}(t_p)=(\gamma_m/\gamma_c)^2 
\propto (\epsilon_e\epsilon_{B,r}\Gamma_d^4t_pn_1)^2$ 
for the magnetization estimate where $\gamma_m$ and $\gamma_c$ are 
the random Lorentz factors of electrons corresponding to the 
typical and cooling break frequencies, respectively.}. 
Finally, in the case (d), the forward shock emission
also peaks at the onset of afterglow, and it follows that 
$R_t=1$. It is possible to show that Equation (\ref{eq:RB}) is still
valid. 

When an early afterglow light curve shows a flattening at $t=t_{flat}$
after the steep decay phase (the bottom panel in figure
\ref{fig:twopeaks}), the reverse shock emission dominates at early
times. The forward shock peak is masked by the reverse shock emission,
the peak ($t_{p,f}$, $F_{p,f}$) is not observationally 
determined. In such a case, the upper limit $t_{p,f}=t_{flat}$ gives a
rough estimate of $R_B$. Considering that the reverse and forward shock emission
components are comparable at the flattening, we obtain $R_F \sim
R_t^{\alpha}$. Substituting this relation into Equation (\ref{eq:RB}), we get 
\begin{equation}
\label{eq:Rb}
R_B \sim \paren{\frac{R_t^3\Gamma_0^{4\alpha-7}}
{C_F^3C_m^{2(\alpha-1)}}}^{2/(2\alpha+1)} 
\sim \paren{\frac{R_t^3\Gamma_0}{C_F^3C_m^2}}^{2/5}.
\end{equation}
where $R_t=t_{flat}/t_{p}$.
If the forward shock emission peaks earlier $t_{p,f}<t_{flat}$, 
the real value of $R_B$ might be slightly different. To evaluate 
how $R_B$ depends on $t_{p,f}$, we refer to the scalings $R_t \propto t_{p,f}$
and $R_F \propto t_{p,f}^{\alpha_f}$ where $\alpha_f$ is the decay index 
of the forward shock emission. Using these scalings, one finds that 
the dependence is weak: 
$R_B \propto t_{p,f}^{6(1+\alpha_f-\alpha)/(1+2\alpha)}$.
If the forward shock decays as the theory suggests 
$\alpha_f=3(p-1)/4$, a relation 1+$\alpha_f-\alpha=0$ holds, 
and $R_B$ does not depend on $t_{p,f}$ (Gomboc et al. 2008). 

For weakly magnetized fireballs, the ratio $\sigma$ between
the Poynting flux energy  and the kinetic energy (the baryonic
component) around $t_p$ is expressed as a function of the magnetization parameter $R_B$ as 
\begin{equation}
\label{eq:sig}
\sigma \sim  \paren{\frac{\bar{\Gamma}_d-1}{\bar{\Gamma}_d}} \epsilon_{B,f} R_{B},
\end{equation}
where $\bar{\Gamma}_d$ is the Lorentz factor of shocked shell material 
relative to the unshocked shell.

\section{Shocks in the intermediate regime}
\label{sec:inter}
The simple estimates of $r_d$ and $\Gamma_d$, which we have discussed in 
section \ref{sec:FSRS}, provide useful insights into the fireball
dynamics. However, these are order-of-magnitude estimates, and 
obtained by assuming that the reverse shock is ultra-relativistic or
Newtonian. Since most observed bursts are actually in the intermediate
regime  $\xi_0 \sim 1$, we here consider a better approximation which is
similar to one discussed by Nakar \& Piran (2004). 

\begin{figure}
\begin{center}
\includegraphics[width=16cm]{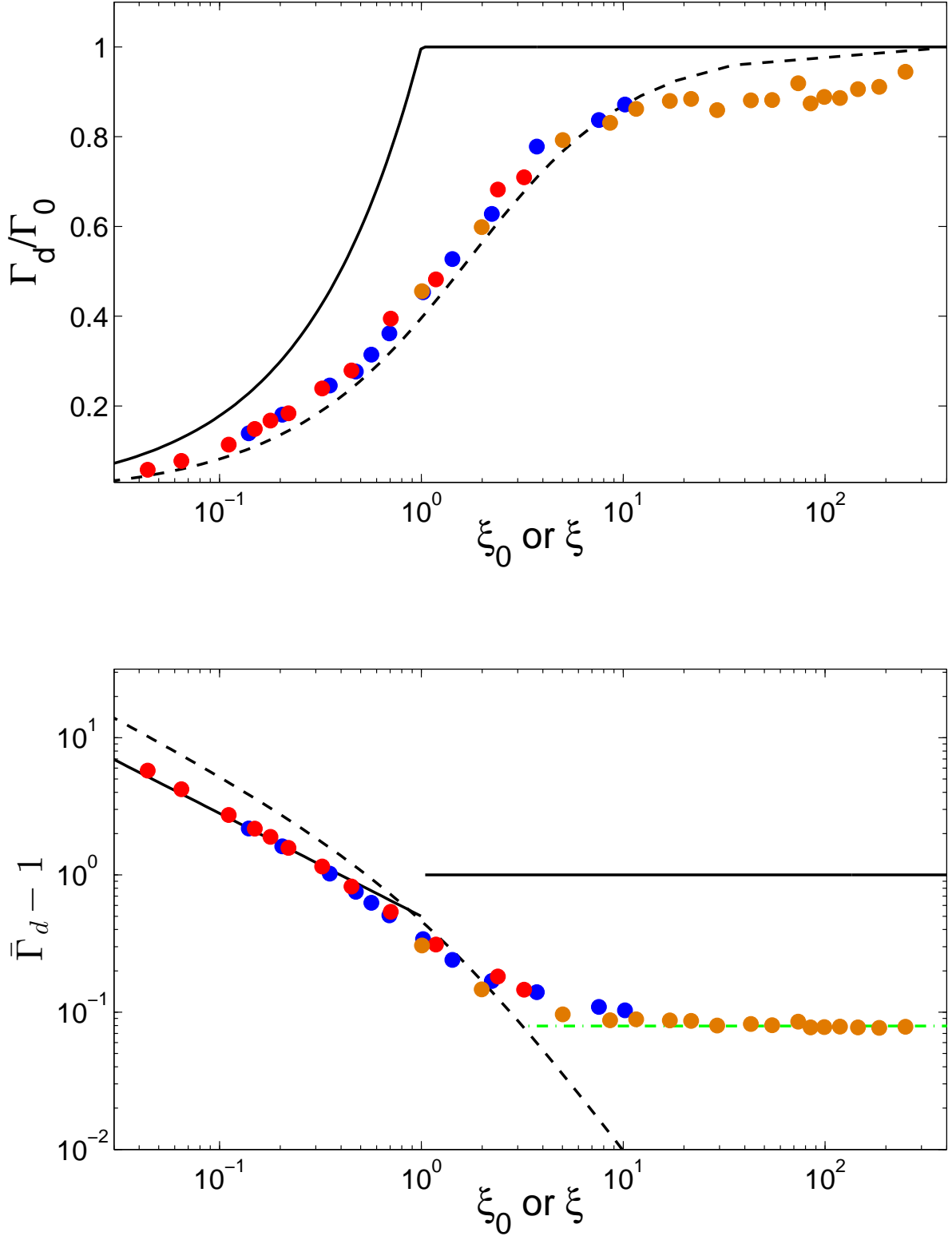}
\caption{ 
$\Gamma_d/\Gamma_0$ (Top panel) and 
$(\bar{\Gamma}_d-1)$ (Bottom panel) as functions of $\xi_0$ or
 $\xi$. The conventional estimates (black solid lines), 
the approximation (\ref{eq:intermediate}) (black dashed lines),
and the numerical results (orange dots: $r_{0}=10^{9}$cm,
blue dots: $6\times10^{11}$cm, and
red dots: $6\times10^{12}$cm).
See the text concerning to the choice of $\xi_0$ or $\xi$.
\label{fig:interm}}
\end{center}
\end{figure}

The deceleration of an expanding shell happens when it gives a
significant fraction of the kinetic energy to the 
ambient medium. Equalizing the energy in the shock ambient matter
with $E/2$, we obtain $r_{d} = 2^{-1/3}l/\Gamma_d^{2/3}$. The Lorentz
factor $\Gamma_d$ in the shock regions is given as a function of the
initial Lorentz factor $\Gamma_0$ and the density ratio $n_4/n_1$
between the  unshocked shell material and the ambient medium (Sari \&
Piran 1995).  For a homogeneous shell with width $\Delta$, the
particle density is   
$n_4 \sim (E/m_pc^2\Gamma_0)/(4\pi r_d^2\Delta \Gamma_0)$. 
Since the shock jump conditions and equality of pressure 
along the contact discontinuity give a relation 
$n_4/n_1 \sim 4\Gamma_d^2/\bra{(4\bar{\Gamma}_d+3)(\bar{\Gamma}_d-1)}$, 
we get an equation 
\footnote{Assuming $r_{d} = l^{3/4}\Delta^{1/4}$, Nakar \& Piran (2004)
have obtained a similar equation.}
for $x\equiv \Gamma_d/\Gamma_0$ as
\begin{equation}
\xi^2 \sim \frac{24x^{8/3}}{2^{2/3}(1-x)^2(2+3x+2x^2)},
\label{eq:intermediate}
\end{equation}
where $\xi=(l/\Delta)^{1/2}\Gamma_0^{-4/3}$ and 
we have used $\bar{\Gamma}_d\sim (x+1/x)/2$. 
The corresponding results are shown in figure \ref{fig:interm}. 
For $\xi \ll 1$, we obtain $x \sim 0.47 \xi^{3/4}$, while for $\xi \gg 1$,
we obtain $x \sim 1$. 
In the rest of the paper, we call the estimates obtained in 
this section as the approximation (\ref{eq:intermediate}) or the
estimates based on Equation (\ref{eq:intermediate}), while
the estimates discussed in section 2 are called the conventional
estimates. Since Equation (\ref{eq:intermediate}) gives 
$x$ for a given $\xi$, to estimate $x$ at the deceleration radius $r_d$,
we need to use 
the value of $\xi$ at $r_d$. In the thick shell regime $\xi_0 <1$, $\xi$ is
a constant during the deceleration process and we can use the initial
value $\xi_0$. However, in the thin shell regime $\xi_0 \gg 1$, due to the shell's
spreading $\Delta\sim\Delta_0+r_d/\Gamma_0^2\sim\Delta_0(1+\xi_0^2)$,
the value $\xi\sim \xi_0 (1+\xi_0^2)^{-1/2}$ is always about unity 
at $r_d$ (Sari \& Piran 1995). 
Then, if we plot $x$ and the reverse shock temperature 
$(\bar{\Gamma}_d-1)$ as functions of the initial value $\xi_0$, they are
expected to flatten in the intermediate regime. Since $\xi\sim \xi_0$ in
the thick shell and intermediate regime, we can directly compare the two
approximations, and we find that the conventional approximation
overestimates $x$ and $(\bar{\Gamma}_d-1)$ in the intermediate regime.
The conventional estimates are
$x = \xi_0^{3/4}$ and $(\bar{\Gamma}_d-1) = \xi_0^{-3/4}/2$ 
for $\xi_0 <1$ and $x\sim 1$  and $(\bar{\Gamma}_d-1) \sim 1$ for $\xi_0 >1$.

Using the deceleration radius $r_d$ and the Lorentz factor $\Gamma_d$,
the deceleration time is $t_d \sim
r_d/(2c\Gamma_d^2)=l/(2^{4/3}c\Gamma_d^{8/3})$. For the solution 
of Equation (\ref{eq:intermediate}), we have an estimate 
of the correction factor $C_t=2^{-4/3}x^{-8/3}$.
Assuming no gradients in the distribution functions of 
the pressure and velocity in the shock regions, we obtain 
$\nu_{m,r}/\nu_{m,f} \sim (\bar{\Gamma}_d-1)^2/\Gamma_d^2$ and 
$F_{\nu,max,r}/F_{\nu,max,f}\sim \Gamma_d^2/\Gamma_0$
where we have assumed $R_B=\epsilon_{B,r}/\epsilon_{B,f}=1$. 
Then, we get the correction factors 
$C_m\sim (1-x)^4/(4x^{4})$ and $C_F\sim x^2$. 

%%%%%%%%%%%%%%%%%%%%%%%%%%%%%%%%%%%%%%%%%%%%%%%%%%%%%%%%%%%%%%%%%%%%%%%%%%%%%%%%%%%%%%%%%%%%%%%%%%%%
\section{Numerical Simulation}
\label{sec:Method}

The two analytic estimates which we have discussed include approximations
(e.g. a simplified shock approximation and no gradients in the
distribution functions of hydrodynamics quantities in the shock regions).
Furthermore, the estimate  (\ref{eq:intermediate}) gives 
the Lorentz factor $\Gamma_d$ for a given $\xi$ at the deceleration
radius $r_d$, instead of the initial value $\xi_0$. Since the typical
frequency of the reverse shock emission is sensitive to the temperature
$(\bar{\Gamma}_d-1)$, it is important to investigate 
how $\xi$ at $r_d$ depends on $\xi_0$ (or where $\xi$ becomes a constant)
and what asymptotic value the reverse shock 
temperature takes in the thin shell regime. 
To examine the accuracy of the
approximations and evaluate the shock Lorentz factors and the
corrections factors $C_t$, $C_m$ and $C_F$, we employ a spherical 
Lagrangian code based on the Godunov method with an exact
Riemann solver (Kobayashi et al. 1999; Kobayashi \& Sari 2000; Kobayashi
\& Zhang 2007). No MHD effects are included in our purely hydrodynamic 
calculations. However, if the magnetization of a fireball is not too
large (i.e. the ratio of magnetic to kinetic energy flux $\sigma \lsim
0.1$; Giannios et al. 2008; Mimica et al. 2009), 
the dynamics of shocks is not affected by 
magnetic fields, and our numerical results can be used
to model the synchrotron emission from forward and reverse shocks.
We will evaluate the correction factors for
$R_B=\epsilon_{B,r}/\epsilon_{B,f}=1$. 

The initial configuration for our simulations is a static uniform
fireball surrounded by a uniform cold ISM. The hydrodynamic evolution is
evaluated through the stages of initial acceleration, coasting, energy
transfer to the ISM and deceleration. The evolution of a fireball is
fully discussed in Kobayashi et al. (1999). We assume explosion energy
$E_{0}=10^{52}$ ergs and ambient density
$n_{1}=1~\mbox{proton~cm}^{-3}$ for all the simulations, while  we vary
the dimensionless entropy ($40~<~\Gamma_{0}~<~10^3$) and  
the initial fireball size ($r_0~=~ 10^{9}$cm, $6\times10^{11}$cm,
or $6\times10^{12}$cm) to cover a wide rage of $\xi_0$.  

\subsection{Spectra and Light Curves}

We evaluate shock emission as a sum of photons from Lagrangian
cells (fluid cells) in numerical calculations. First consider a single fluid
cell with Lorentz factor $\Gamma$, internal energy density $e$, particle
density $n$ and mass $m$ in a shocked region (forward or reverse shock).
Electrons are assumed to be accelerated in the shock to a 
power-law distribution with index $p=2.5$ above a minimum Lorentz factor
$\gamma_m$. We assume that constant fractions $\epsilon_e=6\times10^{-2}$ and
$\epsilon_B=6\times10^{-3}$ of the shock energy are given to electrons and
magnetic fields, respectively. Our results are insensitive 
to the exact values of the microphysical parameters as long as $R_B=1$, but
they are included here for completeness. The typical random Lorentz factor 
and the energy of magnetic fields evolve as $\gamma_m \propto e/n$
and $B^2 \propto e$. The typical synchrotron frequency in the observer
frame is  $\nu_m \propto \Gamma \gamma_m^2B$, and the peak spectral
power is $F_{\nu,max} \propto N_e \Gamma B$ for a total number 
$N_e\sim m/m_p$ of electrons in the cell.
As we use a Lagrangian code, $N_e$ remains constant throughout the
numerical evolution.  The flux at a given frequency above $\nu_m$ is 
$F_\nu = F_{\nu,max}(\nu/\nu_m)^{-(p-1)/2}\propto 
m\Gamma^{(p+1)/2}e^{(5p-3)/4}n^{1-p}$ , while below $\nu_m$ 
we have a synchrotron low-energy tail as 
$F_{\nu}=F_{\nu,max}(\nu/\nu_m)^{1/3} \propto
m\Gamma^{2/3}e^{-1/3}n^{2/3}$. 
Then, the emission from a cell can be estimated by using
hydrodynamic quantities. 

We treat a fluid cell as a particle that continually
emits photons. However, we only have the locations of the cell 
$\{r_j\}$ and flux estimates $\{F_{\nu,j}\}$ at discrete timesteps 
$\{t_{lab,j}\}$ where the subscript $j$ indicates quantities at
lab timestep $j$ and $r_j$ indicates the inner boundary of the cell. 
We assume all the photons are emitted from the inner boundary (i.e. we
neglect the radial width of the cell). Prior to the light curve
construction we generate a series of (logarithmic) bins with 
boundaries $\{t_k\}$ in the observer time domain, and we assume bin $k$
is bounded by $t_k$ and $t_{k+1}$. We now  consider the emission from 
a single fluid cell between two consecutive lab timesteps: 
$j$ and $j+1$. Since a photon emitted at timestep $j$ arrives at
the observer at  $t_{j} \equiv t_{lab,j}-r_j/c$, photons spread over
observer time bins between $t_j$ and $t_{j+1}$. Note that the
observational time $t_j$ monotonically increases with $j$. 
Assuming that the observed flux $F_{\nu}$ evolves linearly between $t_j$ and 
$t_{j+1}$, and that the observer detects photons between $\nu_R$ and
$\nu_R+d\nu$, we can estimate the amount of energy deposited to each
time bin. 

By monitoring the entropy evolution of a fluid cell, we can determine
when the cell is heated by a forward or reverse shock. Then, we take
into account all the timesteps after the shock heating for the
construction of the light curve of the single fluid cell.  We can apply
this technique to all the cells inside (or outside) the contact
discontinuity, and the total energy from all the cells in each time bin
is divided by the bin size to get the reverse (or forward) shock light
curve. It is then the simple matter of finding the maximum flux to
obtain the peak time $t_p$ of the reverse shock emission. 

To numerically define a property $f$ of the fireball shell at the peak time, we
consider an average value   
\begin{equation}
\left \langle f \right \rangle = \frac{\sum f_i \delta E_i}{\sum 
 \delta E_i}.
\label{eq:ave}
\end{equation}
where the summation is taken over all the fluid cells $\{i\}$  which are
inside the contact discontinuity (i.e. in the reverse shock region) and
which have contributed to the peak flux and $\delta E_i$ is the 
contribution from fluid cell $i$ to the peak time bin. 

At the peak time, we construct the
spectrum. For this purpose, we set up a series of bins $\{\nu_q\}$
in the frequency domain.  For the peak time bin (i.e. the time bin which
gives the peak flux), we know which fluid cells have contributed, and at
which lab timestep it has happened. Let us assume that a fluid cell
deposits energy between lab timestep $j$ and $j+1$. 
Assuming a linear evolution of the luminosity
$\int_{\nu_q}^{\nu_{q+1}} F_{\nu} d\nu$ between the timesteps,
we can estimate how much energy is deposited in each frequency bin
at the peak time (the peak time bin). After summing up all the
energy deposited by the relevant fluid cells in each 
frequency bin, we divide the energy by the frequency bin size to get the
spectrum at the peak time. 

\subsection{Comparison of the Estimates and the Correction Factors}
\label{sec:results}

Figure \ref{fig:interm} shows the Lorentz factor $\Gamma_d$ and the
reverse shock temperature $(\bar{\Gamma}_d-1)$ at the peak time
$t_p$.  For the numerical results (the dots), we have used
Equation (\ref{eq:ave}) with $(\bar{\Gamma}_d-1)=e/nm_pc^2$ 
to obtain the average values, and we have assumed $\Delta_0=r_0$ to
estimate $\xi_0$.  The numerical results and the conventional
approximation are plotted against 
$\xi_0$, while the approximation (\ref{eq:intermediate}) 
is plotted against $\xi$. As we have discussed in Section
4, when the initial value is high $\xi_0 \gg 1$, the $\xi$ parameter is
expected to decrease to order-of-unity during the evolution.  One finds
that such flattening in the numerical results occurs at a rather high
value $\xi_0 \sim$ several.  The approximation (\ref{eq:intermediate}),
especially $(\bar{\Gamma}_d-1)$ in the 
intermediate regime and $\Gamma_d$ in almost the whole range, is in
better agreement with the numerical results, compared to the
conventional estimates. The green dashed-dotted line in the bottom panel
indicates the numerical asymptotic value: 
$\left \langle \bar{\Gamma}_d -1 \right \rangle \sim 8\times10^{-2}$.   

\begin{figure}
\begin{center}
\includegraphics[width=16cm]{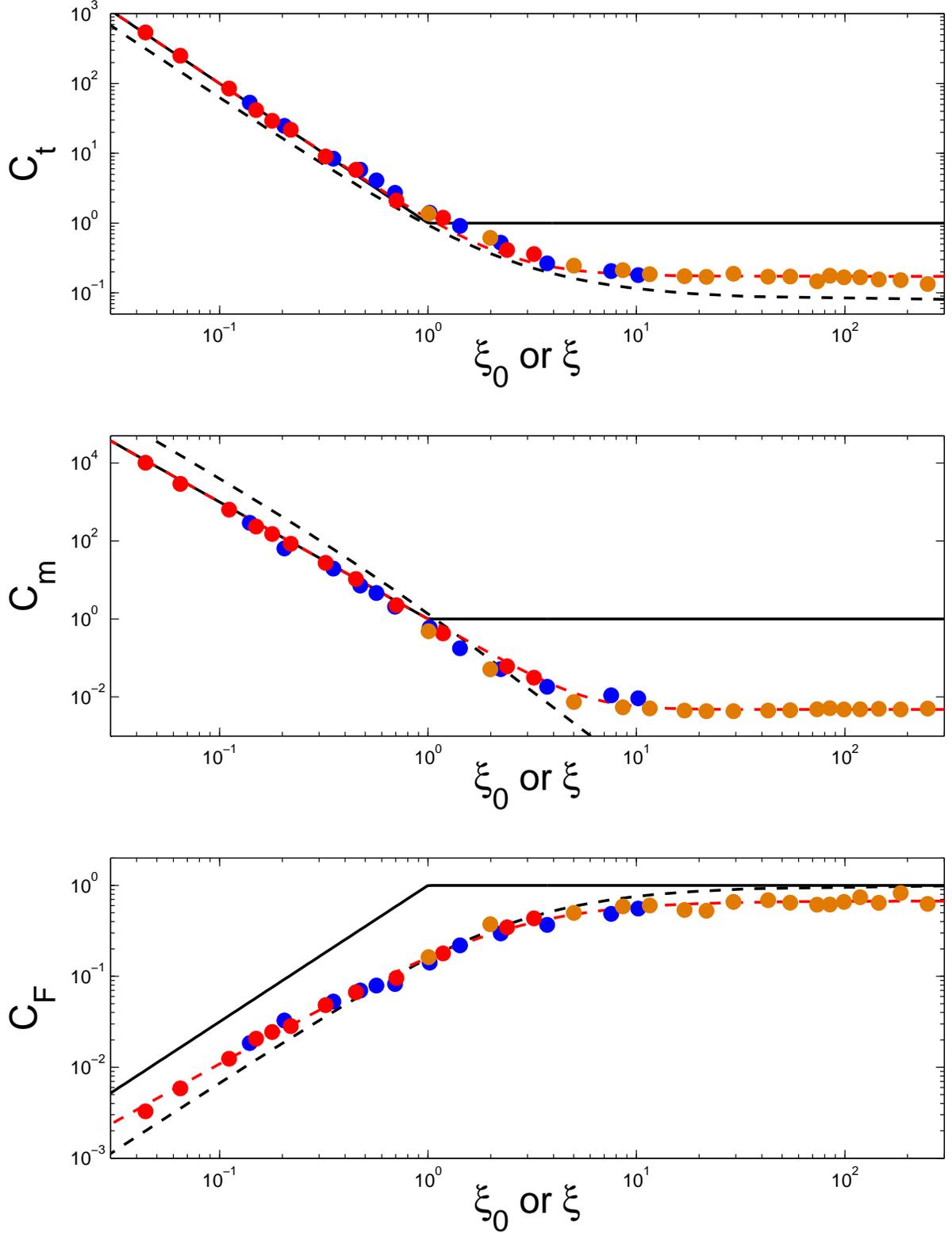}
\caption{The correction factors as a function of $\xi_0$ or $\xi$. 
Top panel: The normalized peak time $C_t$, 
Middle panel: The normalized typical frequency ratio $C_m$, and 
Bottom panel: The normalized peak flux ratio $C_{F}$.
The conventional estimates (black solid lines),
the estimate based on the approximation (\ref{eq:intermediate}) (black
dashed lines), the numerical results (blue, red and orange dots are the same as
 in Figure \ref{fig:interm}), and numerical fitting formulae (red dashed lines).
\label{fig:all}}
\end{center}
\end{figure}

Using  a numerical peak time $t_p$, we estimate the correction
factor $C_t=c t_p \Gamma_0^{8/3}/l$ as a function of $\xi_0$.
The results are shown in the top panel of Figure \ref{fig:all}. 
The conventional approximation well explains the 
numerical results in the thick shell regime $\xi_0<1$ 
but it breaks down in the thin shell regime $\xi_0>1$. 
In the thin shell regime, the numerical $C_{t}$ is lower by a factor 
of $\sim 5$ than the conventional estimate which is equivalent to the
numerical peak time being earlier than expected. 
Since for simplicity we have neglected a factor of 2 in the
conventional estimate as $t_d=l/c\Gamma_0^{8/3}$
instead of $l/2c\Gamma_0^{8/3}$, $C_t=0.5$ would be more appropriate 
for the conventional estimate in the thin shell regime. However, 
the numerical results are still smaller. This is in part due to the
gradients in the distribution functions of hydrodynamics quantities in the
reverse shock region. The numerical distributions have a higher value at
the contact discontinuity, and  they decrease toward the tail (see Figure 3 in
Kobayashi \& Zhang 2007). It makes the contribution of photons from 
the inner parts of the fireball shell less significant, reducing
the effective width of the emission region in the shell, and the
shock  emission peaks earlier than in the homogeneous shell case. 
Since as we will see later, the magnetization estimate is rather
insensitive to $C_t$ (and the peak time), we discuss only the line-of-sight emission in
this paper. However, it is possible to include the high latitude
emission at expense of computational time, and we have obtained very
similar results for several selected cases. With the addition of 
the high latitude emission the overall light
curve appears smoother with slightly shallower decay features. The
position of the peak time increases by $\sim 50\%$. 

Using the numerical values of the typical frequency ratio at the peak 
time, we estimate the correction factor 
$C_m=\Gamma_0^2 \bra{\nu_{m,r}(t_p)/\nu_{m,f}(t_p)}$ as a function of 
$\xi_0$. The results are shown in the middle panel of Figure \ref{fig:all}.
The conventional estimate is in good agreement  with the numerical
results in the thick shell regime, but as we expect from  
$C_{m}\varpropto (\bar{\Gamma}_{d}-1)^{2}$, it overestimates $C_m$ 
by a factor of $\sim 10^2$ in the thin shell regime. Finally the bottom
panel of Figure \ref{fig:all} shows the results for 
$C_F=\Gamma_0^{-1}\bra{F_{\nu,max,r}(t_p)/F_{\nu,max,f}(t_p)}$.
The conventional approximation overestimates the amount of flux emitted
by the reverse shock especially in the intermediate regime as 
Nakar \& Piran (2004) have pointed out. 
The estimates based on Equation (\ref{eq:intermediate}) 
provide a better approximation for all three correction factors
in the intermediate regime. The red dashed lines in the 
three panels indicate
the numerical fitting formulae $C_{t}=N_t+\xi_0^{-2}$ with $N_{t} \sim
0.2$, $C_{m}=N_{m}+\xi_0^{-3}$ with $N_{m} \sim 5\times10^{-3}$ and 
$C_F^{-1}=N_{F}+M_{F}\xi_0^{-P_{F}}$ with 
$N_F\sim1.5$, $M_F\sim5$ and $P_F\sim1.3$. 

\begin{figure}
\begin{center}
\includegraphics[width=16cm]{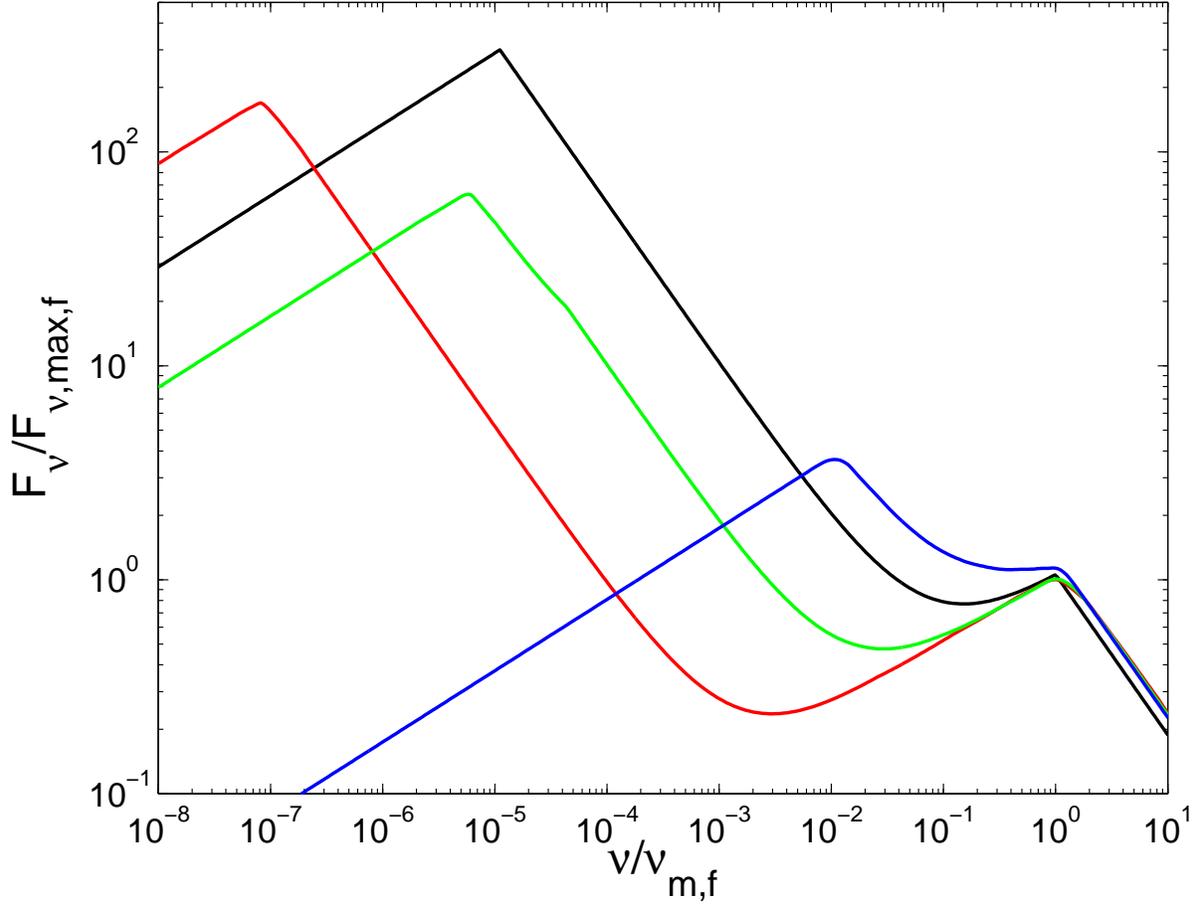}
\caption{ Numerical wide band spectra at the peak time:
$\xi_0=0.1$ (blue line), 1 (green line) and 10 (red line).  
The black line shows the conventional estimate (thin shell case).
The frequency and flux are normalized by the typical frequency and 
peak flux of the forward shock emission, respectively.
$\Gamma_{0}=300$ and $\Delta_0\sim 3\times 10^9$cm (red line), 
$\sim 3\times 10^{11}$cm (green line),
or $\sim 3\times 10^{13}$cm (blue line) are assumed. 
\label{fig:spec}}
\end{center}
\end{figure}

Figure \ref{fig:spec} illustrates wide band spectra at the peak time.
We here consider three numerical cases with $\xi_0=0.1, 1$ or 10.
The black line indicates the conventional estimate in which
the typical frequency $\nu_{m,r}$ of the reverse shock
emission is lower by a factor of $\Gamma_0^2$ than that of the forward
shock emission, and the peak  flux $F_{\nu,max,r}$ of the reverse shock
emission is higher by a factor of $\Gamma_0$. However, our numerical
results show that in the thin shell regime $\nu_{m,r}$ is lower 
by a further factor of $\sim 10^{2}$ than the conventional estimate (the
red line)
\footnote{If the typical frequency $\nu_{m,r}$ is as low as $\sim 10^{12}$Hz,
the spectrum of the reverse shock emission would peak at the synchrotron 
self-absorption frequency (Nakar \& Piran 2004).}, 
and that in the intermediate regime the peak flux
$F_{\nu,max,r}$ would be lower by a factor of several (the green line).  These indicate that
the reverse shock emission would be elusive if the typical frequency of
the forward shock is around the optical band and if the forward and
reverse shock have the same  microscopic parameters 
(Nakar \& Piran 2004; Mimica et al. 2010; Melandri et al. 2010). 
In the thick shell regime, the peak 
frequency of the reverse shock emission is closer to that  
of the forward shock emission $\nu_{m,r}/\nu_{m,f}\sim \xi_0^{-3}\Gamma_0^{-2}$
(the blue line). We might have a better chance to
detect the reverse shock component in early afterglow,  although
the light curve peaks earlier than in the thin shell regime.  

\subsection{Initial Lorentz Factor and  Magnetization Parameter}

The initial Lorentz factor can be evaluated by using the peak time
$t_{p} \sim t_d$,  
\begin{equation}
\Gamma_0=\paren{\frac{C_tl}{c t_{p}}}^{3/8}
\label{eq:g0}
\end{equation}
where $l$ is basically a known quantity from the prompt gamma-ray and
late-time afterglow observations, and the estimate depends very weakly
on more fundamental parameters $\Gamma_0\propto l^{3/8}\propto 
(E/n_1)^{1/8}$, and we had obtained numerical result $C_t \sim 0.2+\xi_0^{-2}$.
In principle, we can estimate $\xi_0$ from observable
quantities. Since the duration $T$ of the prompt 
gamma-rays gives a rough estimate of the width $\Delta_0 \sim cT$
(Kobayashi et al. 1997), using Equation (\ref{eq:g0}) and the numerical 
$C_t(\xi_0)$, we obtain $\xi_0\sim 5^{1/2} \sqrt{(t_p/T)-1}$ and 
$C_t\sim 0.2 (1-T/t_p)^{-1}$. In the thin shell regime, an early 
afterglow peaks well after the prompt gamma-ray emission (Sari 1997), 
and we have $C_t \sim 0.2$. However, in the thick shell regime, the peak time
is almost equal to the width $\Delta_0/c$. The approximation
$\Delta_0= cT$ might not be accurate enough to discuss the exact value
of $\xi_0$. Since the flux before the optical peak $t_p$ is sensitive to
the initial profile of the shell and in particular to $\xi_0$,
the rising index of the light curve might be used to break the 
degeneracy of the $\xi_0$ estimate in the thick shell regime. 
Nakar \& Piran (2004) have numerically estimated the rising index for a
homogeneous shell in a range of $0.05 < \xi_0 <5$ as 
$\alpha_{rise} \sim 0.6\bra{1+p(\xi_0-0.07\xi_0^2)}$. A slow (rapid) rise is
a signature of the thick (thin) shell regime (Kobayashi 2000).

\begin{figure}
\begin{center}
\includegraphics[width=16cm]{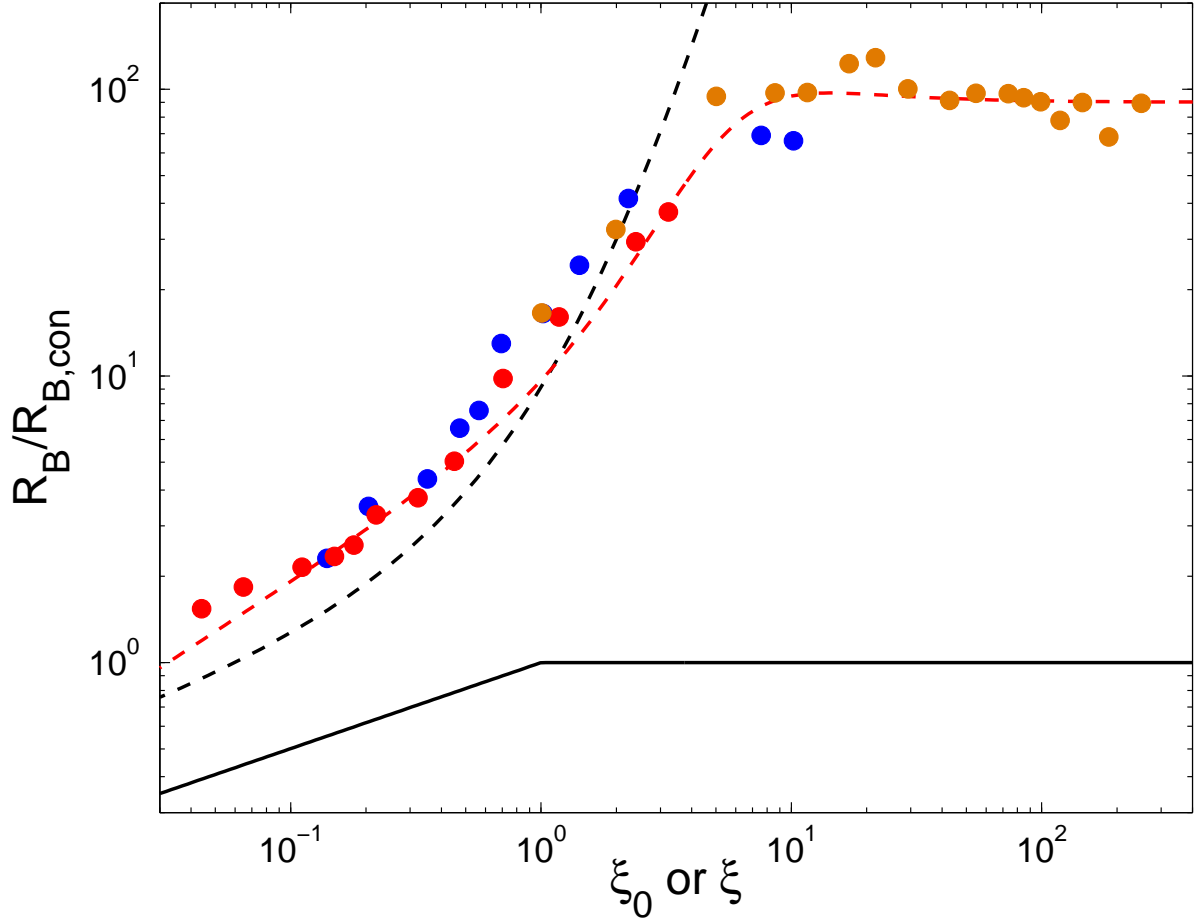}
\caption{The correction factor for the magnetization parameter
 $R_{B}/R_{B,con}$ as a function of 
 $\xi_0$ or $\xi$. The legend is the same as figure
 \ref{fig:all} with the best fit equation (red dashed line) 
 obtained by the combination of the best fits
 to $C_t, C_m$ and $C_F$.  $\alpha=2$ is assumed. 
\label{fig:corr}}
\end{center}
\end{figure}

The magnetization parameter $R_B$ can be estimated by using $\Gamma_0$
$R_{t}$, and $R_{F}$. For the typical decay index of
the reverse shock emission $\alpha=2$, the conventional estimate 
is $R_{B,con}=(R_F^3\Gamma_0/R_t^3)^{2/5}$ 
in the thin shell regime where $\Gamma_0=(l/ct_p)^{3/8}$.
Then, we obtain a correction factor for the 
magnetization parameter as 
\begin{equation}
\frac{R_B}{R_{B,con}}=\frac{C_t^{3/20}}{C_F^{6/5}C_m^{4/5}}.
\label{eq:con}
\end{equation}
One finds that the estimate is rather insensitive to 
$C_t$. In Figure \ref{fig:corr}, the numerical results are plotted 
with the approximations.  For a typical GRB ($\xi_0\sim1$),
the conventional approximation (black solid line) underestimates the
magnetization parameter by a factor of $\sim 10$. A more extreme
discrepancy occurs in the thin shell regime, and the magnetization
parameter is underestimated by a factor of $\sim 10^2$.   
The estimate based on Equation (\ref{eq:intermediate}) (black dashed
line) describes the numerical results reasonably well in the intermediate 
and thick shell regime. 

\section{Case Studies}
\label{sec:case}

\begin{figure}
\begin{center}
\includegraphics[width=16cm]{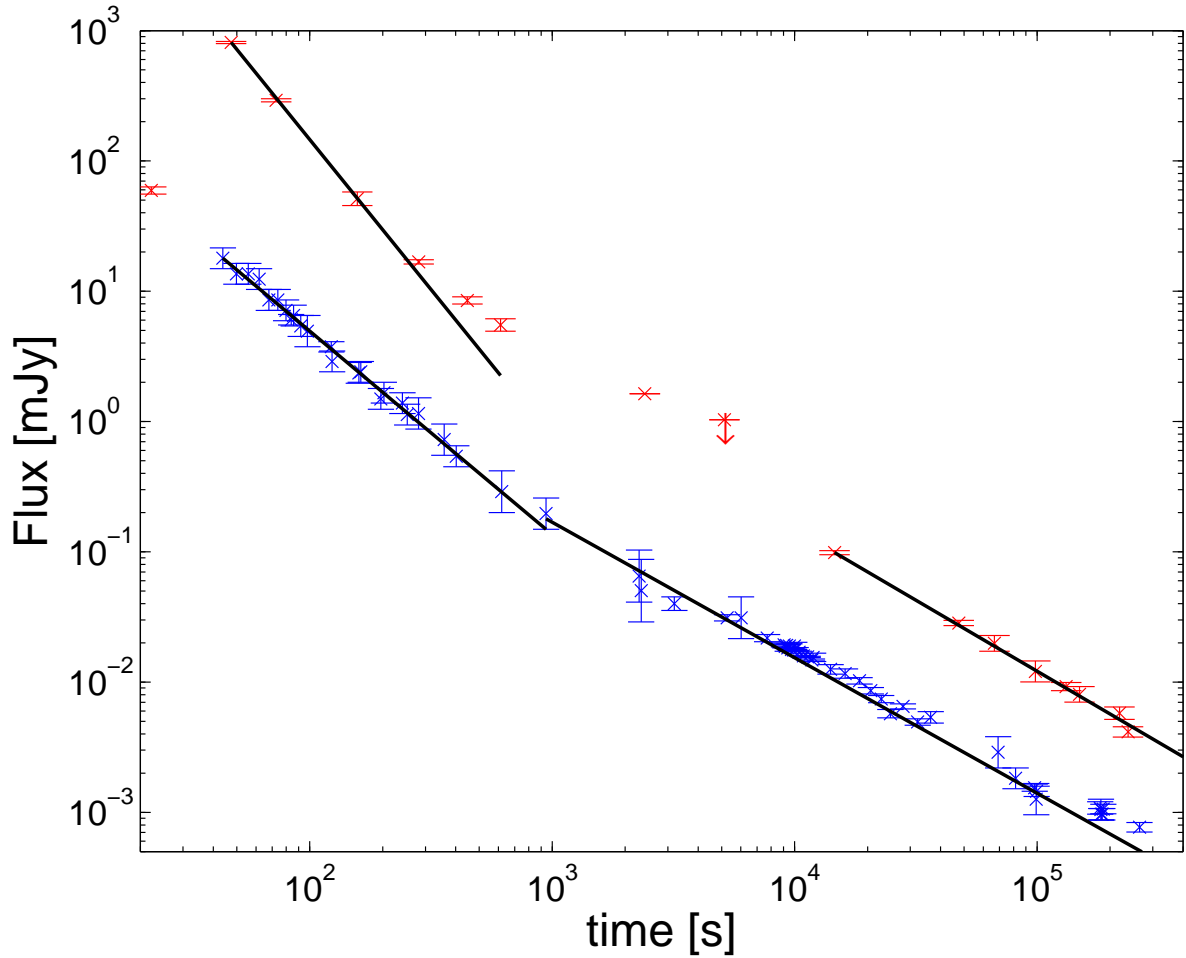}
\caption{Light curves of early optical afterglows: GRB 990123 (red points) and GRB 090102
 (blue points). The solid lines depict power-law fitting to the
 forward shock and reverse shock emission components. 
 Data are from Kulkarni et al. 1999; Kobayashi \& Sari 2000; Gendre et
 al. 2010 and references therein. 
\label{fig:lc}}
\end{center}
\end{figure}

Swift revealed that the behavior of early afterglows are complicated
than expected and there are indications of long-lasting central 
engine activity (e.g. flares and late-time energy injection;
Zhang et al. 2006; Nousek et al. 2006; O'Brien et al. 2006). Although
the nature of the central engine activity and additional components in
early afterglows are interesting research subjects, in order to
demonstrate our scheme which is based on an impulsive explosion model,
we discuss early optical afterglows associated with GRB 990123 and GRB
090102. These light curves are described by a broken power-law with no
flares (see Figure \ref{fig:lc})

{\bf GRB 990123:}
This burst is one of the brightest GRBs observed so far. The basic
parameters include $z=1.6$, $E\sim1.4\times10^{54}$ergs, and $T\sim 60$s
(e.g. Kobayashi \& Sari 2000 and references therein).
The gamma-ray profile is dominated by two pulses, each lasting $\sim8$s,
separated by 12s. A bright optical flash is detected during the prompt 
emission (Akerlof et al. 1999), the optical emission peaks at $t_p\sim
50$s at 9 mag, and initially rapidly decays and it becomes shallower at
a late time $t_{p,f} < 0.1$ days. Using the bootstrap method for the light
curve fitting, we find that $\alpha=2.31\pm 0.38$ and
$\alpha_f=1.09\pm 0.07$ where the errors quoted are values to within 3
$\sigma$ of the best fits. 
We have only one optical data point before the peak, and it 
provides a lower limit of the rising index $\alpha_{rise}>2$ and the peak flux. 
We conservatively assume that the peak flux is 9 mag.
Since the optical peak is comparable to the GRB duration (especially to the 
duration of the main two pulses) and the rising is rapid, this is an
intermediate case $\xi_0\sim1$, with the correction factors $C_t\sim1.2
$, $C_m\sim1 $ and $C_F \sim0.16 $. Using Equation (\ref{eq:g0}) with a
time-dilation correction, we obtain the initial Lorentz factor is about
$\Gamma_0 \sim 460n_1^{-1/8}$. Assuming $t_{p,f} = 0.1$days, one has  
$R_{t} \sim 170$ and $R_{F} \sim 5000$, Equation (\ref{eq:RB}) gives
the magnetization parameter $R_B \sim 6300^{6600}_{6000}$ where the
subscript and superscript indicate the range of the value when
the error in $\alpha$ is taken into account. 
Since the forward shock peak is masked by
the reverse shock emission, the peak time $t_{p,f}$ is rather
uncertain. As we have discussed in Section 3, the
magnetization estimate depends on $t_{p,f}$ as 
$R_B \propto t_{p,f}^{6(1+\alpha_f-\alpha)/(1+2\alpha)} \sim
t_{p,f}^{-0.23}$. If the
forward shock emission also peaks at $t_{p,f}\sim 50$s, the
magnetization parameter $R_B$ is larger by a factor of $\sim 3$.
Since Zhang et al. (2003) found a magnetization parameter of
$R_{B}\sim15^2=225$ based on the conventional approximation with
$\xi_0\sim1$, our magnetization estimate is 
larger by more than an order of magnitude. 

{\bf GRB 090102:}
This burst shows a significant polarization at the $10\%$ level in the
early optical afterglow and it suggests a magnetized fireball (Steele et al. 2009). 
The basic parameters include $z=1.5$, and $E\sim5.8\times10^{53}$ergs
(e.g. Gendre et al. 2010 and references therein). The prompt gamma-ray 
emission lasts for $27$s and comprises four overlapping peaks 
starting 14s before the GRB trigger. The optical light curve, beginning
at 13 mag, 40s after the GRB trigger, can be fitted by a broken
power-law whose flux decays as a function of time ($F\propto
t^{-\alpha}$) with a gradient $\alpha=1.56\pm0.06$ that then flattens to 
$\alpha_f=1.04\pm0.09$ (the solid lines; a break time is assumed to be 
$10^3$s). If we assume that the
optical emission peaks at the first data point (the mid time is $t_p\sim
60$s after the beginning of the GRB) and $t_{p,f}= 10^3$s, we obtain
$\xi_0\sim 2.4$ and the correction factors: 
$C_t\sim0.3, C_m\sim8\times10^{-2}$, and $C_F\sim0.3$. Using 
 $R_{t}\sim 17, R_{F} \sim 91$ and $\Gamma_0 \sim 230n_1^{-1/8}$, 
we obtain $R_B\sim 220^{310}_{150}$. Since the optical emission has
already decayed at the beginning of the observations, the actual 
peak time $t_p$ might be earlier. If we assume that it peaks at the end
of the prompt gamma-ray emission $t_p\sim 30$s, it would be in the intermediate regime 
$\xi_0 \sim 1$ and $\Gamma_0\sim 290n_1^{-1/8}$. Assuming $t_{p,f}=
10^3$s, we obtain $R_B\sim 140^{180}_{110}$.
The magnetization estimates depend on $t_{p,f}$ as  $R_B \propto t_{p,f}^{0.7}$. 
If $t_{p,f}= 10^2$s, $R_B$ is smaller by a factor of $\sim 5$.

{\bf The $\sigma$ parameter:}
The broadband afterglow emission of GRB 990123 is modeled to find
$\epsilon_{B,f}\sim 10^{-6}$ (Panaitescu \& Kumar 2004). Although
there are no estimates available for GRB 090102, the broadband modeling
generally shows that it is in a range of
$\epsilon_{B,f}=10^{-5}-10^{-1}$ (Panaitescu \& Kumar 2002). Using the 
estimated value of the magnetization parameter $R_B$, the ratio of
magnetic to kinetic energy flux is $\sigma \sim 
10^{-3}(g/0.25)(R_B/6300)(\epsilon_{B,f}/10^{-6})$ 
for GRB 990123 where $g\equiv(\bar{\Gamma}_d-1)/\bar{\Gamma}_d \sim
0.25$ for $\xi_0\sim1$. For GRB 090102, 
assuming $g\sim 0.15 ~(\xi_0\sim2.4)$ and $R_B\sim220$,
it is in a range of $\sigma \sim 3\times10^{-4}-3$.
Although magnetic pressure would suppress the formation of a reverse shock
if $\sigma \simgeq0.1$ (Giannios et al. 2008; Mimica et
al. 2009), the low values are consistent (or the result includes the parameter
region consistent) with the basic assumption in our analysis (i.e. magnetic
fields do not affect the reverse shock dynamics). If a future event indicates
a high value $\sigma \simgeq0.1$, an interesting possibility
to reconcile the problem is that the prompt optical emission (and prompt
gamma-rays) would be produced through a dissipative MHD processes rather than
shocks (Giannios \& Spruit 2006; Lyutikov 2006; Giannios 2008; Zhang \&
Yan 2011). 

Our magnetization estimates are slightly lowered if the blast wave
radiates away a significant fraction of the energy in the early
afterglow $E\propto t^{-17\epsilon_e/16}$ (Sari 1997).  For
$\epsilon_e=0.1$, the blast wave energy becomes smaller by a factor of
$1.7$ between $t=50$s and $0.1$days, by a factor of $1.3$ between
$t=60$s and $10^{3}$s, then the estimates of $R_B$ and $\sigma$ are
reduced by a factor of $1.5-2$.

%%%%%%%%%%%%%%%%%%%%%%%%%%%%%%%%%%%%%%%%%%%%%%%%%%%%%%%%%%%%%%%%%%%%%%%%%%%%%%%%%%%%%%%%%%%%%%%%%%%%
\section{Conclusions}
\label{sec:conclusion}

We have discussed a revised method to estimate the magnetization degree
of a GRB fireball. We use the ratios of observed properties of early 
afterglow so the poorly known parameters for the shock
microphysics (e.g. $\epsilon_{e}$ and $p$) would cancel out. Since the 
estimate depends only weakly on the explosion energy and the fireball
deceleration time, the estimate does not require the exact distance 
(redshift) to the source as an input parameter.
Since most observed events fall in the intermediate regime between the
thin and thick shell extremes, we have provided a new approximation for
the spectral properties of the forward shock and reverse shock 
emission, which well describes the numerical results in
the intermediate regime. The previous standard approach
underestimates the degree of fireball magnetization $R_B$ by a factor of
$10\sim100$. We have estimated $\sigma \sim 10^{-3}$ for GRB 990123. 
For GRB 090102, it is not well constrained due to the uncertainty in
$\epsilon_{B,f}$, and it is in a range of $\sigma \sim 3\times10^{-4}-3$.

In the GRB phenomena, extreme relativistic motion with $\Gamma_0 > 100$
is necessary to avoid the attenuation of hard gamma-rays. The
acceleration process is likely to induce a small velocity dispersion inside
the outflow $\Delta \Gamma \sim \Gamma$ (e.g. thermal acceleration).
If internal shocks are responsible for the production of the prompt
gamma-rays, the dispersion should be even larger (Beloborodov 2000;
Kobayashi \& Sari 2001). The velocity dispersion leads to the spreading
of the shell structure in the coasting phase and 
the $\xi$ parameter decreases as $\xi\propto \Delta^{-1/2}$. As an 
order-of-magnitude estimate, when the initial value $\xi_0>1$, the reverse
shock always becomes mildly relativistic $(\xi \sim1)$ at the
deceleration radius and the reverse shock
temperature $(\bar{\Gamma}_d-1)$ is expected to be insensitive to
the initial value $\xi_0$. However, it is difficult to  analytically
quantify the asymptotic reverse shock temperature.  We have numerically
shown that the spreading effect becomes significant at rather high values
$\xi_0\gsim$ several, and that the asymptotic  value is
$(\bar{\Gamma}_d-1)\sim 8\times 10^{-2}$. 

We have confirmed that, especially in the intermediate regime $\xi_0\sim
1$, the reverse shock emission is much weaker than the standard 
estimates as Nakar \& Piran (2004) had pointed out, 
and that in the thin shell regime the typical frequency of
the reverse shock  emission is much lower than the standard estimates.
If the fireball is not magnetized
$R_B=\epsilon_{B,r}/\epsilon_{B,f}=1$, the reverse shock emission more 
easily falls below the forward shock emission. The lack of optical
flashes from most GRBs might be partially explained in a revised
non-magnetized model. If the fireball shell does not spread 
even in the thin shell regime (i.e. the velocity distribution is completely
uniform), only a small fraction of the kinetic energy of the shell is
converted to thermal energy in the reverse shock,
and the reverse shock emission is practically suppressed in the thin
shell regime $\xi_0 \gg 1$.  

We thank Ehud Nakar and Elena M. Rossi for useful discussion. This research was supported
by STFC fellowship and grant. 

%%%%%%%%%%%%%%%%%%%%%%%%%%%%%%%%%%%%%%%%%%%%%%%%%%%%%%%%%%%%%%%%%%%%%%%%%%%%%%%%%%%%%%%%%%%%%%%%%%%%

%%%%%%%%%%%%%%%%%%%%%%%%%%%%%%%%%%%%%%%%%%%%%%%%%%%%%%%%%%%%%%%%%%%%%%%%%%%%%%%%%%%%%%%%%
%%%%%%%%%%%%%%%%%%%%%%%%%%%%%%%%%%%%%%%%%%%%%%%%%%%%%%%%%%%%%%%%%%%%%%%%%%%
%%%%%%%%%%%%%%%%%%%%%%%%%%%%%%%%%%%%%%%%%%%%%%%%%%%%%%%%%%%%%%%%%%%%%%%%%%%
%%%%%%%%%%%%%%%%%%%%%%%%%%%%%%%%%%%%%%%%%%%%%%%%%%%%%%%%%%%%%%%%%%%%%%%%%%%
%%%%%%%%%%%%%%%%%%%%%%%%%%%%%%%%%%%%%%%%%%%%%%%%%%%%%%%%%%%%%%%%%%%%%%%%%%%
%%%%%%%%%%%%%%%%%%%%%%%%%%%%%%%%%%%%%%%%%%%%%%%%%%%%%%%%%%%%%%%%%%%%%%%%%%%
%%%%%%%%%%%%%%%%%%%%%%%%%%%%%%%%%%%%%%%%%%%%%%%%%%%%%%%%%%%%%%%%%%%%%%%%%%%
%%%%%%%%%%%%%%%%%%%%%%%%%%%%%%%%%%%%%%%%%%%%%%%%%%%%%%%%%%%%%%%%%%%%%%%%%%%
%%%%%%%%%%%%%%%%%%%%%%%%%%%%%%%%%%%%%%%%%%%%%%%%%%%%%%%%%%%%%%%%%%%%%%%%%%%
%%%%%%%%%%%%%%%%%%%%%%%%%%%%%%%%%%%%%%%%%%%%%%%%%%%%%%%%%%%%%%%%%%%%%%%%%%%
%%%%%%%%%%%%%%%%%%%%%%%%%%%%%%%%%%%%%%%%%%%%%%%%%%%%%%%%%%%%%%%%%%%%%%%%%%%
%%%%%%%%%%%%%%%%%%%%%%%%%%%%%%%%%%%%%%%%%%%%%%%%%%%%%%%%%%%%%%%%%%%%%%%%%%%
%%%%%%%%%%%%%%%%%%%%%%%%%%%%%%%%%%%%%%%%%%%%%%%%%%%%%%%%%%%%%%%%%%%%%%%%%%%%%%%%%%%%%%%%%
\end{document}